\documentclass[12pt]{article}

\newtheorem{thm}{Th\'eor\`eme}[section]
\newtheorem{prop}{Proposition}[section]
\newtheorem{lem}{Lemme}[section]
\newtheorem{rem}{Remarque}[section]
\newtheorem{defin}{D\'efinition}[section]

\begin{document}

\title{\Large \bf Remarques sur une m\'ethode de r\'ealisation de
big\`ebres,\\et
alg\`ebres de Hopf associ\'ees \`a certaines r\'ealisations}

\author{Eric Mourre}
\date{\ 2003}

\maketitle



\begin{abstract}
This article introduces a method, which starting from simple and quite general
 mathematical data, allows to construct linear algebras of operators which are, 
 each of them, endowed with a bialgebra structure (coproduct and counity ),
(thm 2.1, 2.2, 2.3) . Moreover under some explicit and natural conditions on theses
mathematical data we obtain linear algebras of operators with the following property:
each of them, is either a Hopf algebra, or its bialgebra structure determines a more
 abstract Hopf algebra associated with it. Finally, we describe a more general abstract
condition for theses bialgebras to admit a unique associated Hopf algebra.\\
The presentation is adapted to the cases where the algebras of linear operators 
 are not finitely generated.\\
This article is restricted to the exposition of the method of construction
and to the proofs of existence and uniqueness of the structures associated with each 
algebra of linear operators that are constructed. \\

Nevertheless, it may be noticed that the ideals of relations associated with 
bialgebras that are obtained determine an algebraic domain which is of theoretical
interest .\\

-----

L'objet de cet article est la pr\'esentation d'une m\'ethode  qui permet \`a
partir
de donn\'ees simples de r\'ealiser des familles d'alg\`ebres d'op\'erateurs
lin\'eaires qui
sont munies d'une structure de big\`ebre   (th\'eo\-r\`emes 2.1, 2.2, 2.3 ).
On met en \'evidence  des r\'ealisations suffisamment g\'en\'erales d'alg\`ebres
d'op\'erateurs lin\'eaires pour lesquelles la structure de big\`ebre
obtenue d\'etermine un id\'eal minimal qui soit aussi un coid\'eal et tel que
la big\`ebre quotient soit une alg\`ebre de Hopf (th\'eor\`emes 3.1, 3.2 ).
Finalement le th\'eor\`eme (4.1), plus abstrait, donne pour une r\'ealisation de
big\`ebre  g\'en\'erale, une condition suffisante, pour que la big\`ebre
admette une alg\`ebre de Hopf canoniquement associ\'ee.
On se limite dans cet article \`a la description de la m\'ethode et \`a la
d\'emonstration de
l'existence et de l'unicit\'e des structures construites sur chacune des
alg\`ebres d'op\'erateurs r\'ealis\'ees.
Les big\`ebres sont obtenues par l'action et en particulier
la repr\'esentation de big\`ebres libres sur des alg\`ebres tensorielles ;
l'alg\`ebre image obtenue par une re\-pr\'esen\-ta\-tion n'est, en g\'en\'eral, pas une
big\`ebre.
Notons que
les id\'eaux des relations,  associ\'es par la construction propos\'ee sont
alors des objets alg\'ebriques, th\'eoriquement int\'eressants.\\  

\end{abstract}

\bigskip\bigskip\bigskip
\textbf{Mots clef:}  R\'ealisation de big\`ebres, alg\`ebres de hopf associ\'ees

Pr\'etirage: CPT-2003/P.4544

\vspace{2 cm}

{ \bf\large Introduction.}
\vspace{0.3 cm}

L'\'etude des big\`ebres et des alg\`ebres de Hopf s'est d\'evelopp\'ee dans le
cadre des groupes classiques et plus r\'ecemment autour des exemples que constituent
les groupes quantiques .\\
Les r\'esultats pr\'esent\'es dans cet article concernent plut\^ot la th\'eorie
g\'en\'erale des big\`ebres, parce que la construction pr\'esent\'ee nous donne des
alg\`ebres d'op\'erateurs lin\'eaires, explicites qui sont chacune munies d'une structure 
de big\`ebre; la construction propos\'ee  pr\'esente en ce sens un int\'er\^et
th\'eorique .\\

La pr\'esentation de la m\'ethode est adapt\'ee \`a la r\'ealisation de big\`ebres 
qui ne soient pas n\'ecessairement finiment engendr\'ees.\\

Dans le paragraphe (1), on introduit les d\'efinitions des diff\'erents types de
cog\`ebres sur lesquelles repose la construction , ainsi que les op\'erateurs invariants
\`a droite qui leurs sont attach\'es. On introduit aussi les big\`ebres tensorielles
contruites sur ces cog\`ebres, dont nous aurons besoin dans la suite .\\

 Le paragraphe (2)  expose la m\'ethode de r\'ealisation des big\`ebres, bas\'ee sur 
le lemme 2.1 et le th\'eor\`eme 2.1; l'existence d'une structure effective de big\`ebre
 est d\'emontr\'ee dans les th\'eor\`emes 2.2 et 2.3 .  Ce dernier th\'eor\`eme 
donne des conditions suffisantes qui permettent d'obtenir des big\`ebres dont l'id\'eal
 des relations est engendr\'e par une r\'eunion d\'enombrable d'espaces vectoriels (de
relations ), qui sont des coid\'eaux de dimensions finies .\\

Le paragraphe (3) d\'ecrit des conditions explicites sur les donn\'ees initiales qui
 permettent de construire des alg\`ebres d'op\'erateurs lin\'eaires ayant chacune la 
propri\'et\'e suivante : soit c'est une alg\`ebre de Hopf, soit la structure de big\`ebre
 dont elle est munie, d\'etermine une alg\`ebre de Hopf, plus abstraite, qui lui est
associ\'ee .\\

Le paragraphe (4), pour des big\`ebres obtenues essentiellement sous les hypoth\`eses 
du th\'eor\`eme 2.3, met en \'evidence, dans un cadre plus g\'en\'eral que celui du
paragraphe (3) , l'hypoth\`ese qui permet d'associer une alg\`ebre de Hopf \`a certaines des
big\`ebres r\'ealis\'ees .

\section{ Quelques \'el\'ements n\'ecessaires \`a  la r\'ealisation
explicite de big\`ebres .}

\vspace{0.3cm}

 Ce paragraphe pr\'esente  les outils \'el\'ementaires qui permettent la
construction
explicite et simple de big\`ebres .\\

En particulier les big\`ebres libres, sont un r\'eceptacle , et les big\`ebres
construites seront des sous alg\`ebres de l'alg\`ebre des op\'erateurs
invariants \`a droite agissant sur une big\`ebre libre.
D'autre par la structure de cog\`ebre  joue un r\^ole particulier dans la
construction,
ainsi que les structures duale et pr\'eduale, pour lesquelles on d\'ecrit
quelques relations
\'el\'ementaires .\\

On introduit ici les d\'efinitions essentielles que l'on utilisera .\\

Les alg\`ebres que l'on consid\`ere sont des alg\`ebres associatives sur
$C$ , avec
unit\'e. Les morphismes d'alg\`ebres pr\'eserveront l'identit\'e .\\

\begin{defin}[Cog\`ebre .]\hspace{1cm}\\
\label{def1}

Une cog\`ebre est un espace vectoriel $V$ muni d'un coproduit  $\Delta $ ,
et d'une
counit\'e   $\epsilon$ ,  qui sont des applications lin\'eaires :
$$ \Delta : V \rightarrow V \otimes V$$
$$ \epsilon :V \rightarrow C  .$$
$\Delta $ est coassociatif:
$$(id\otimes\Delta)\circ \Delta = (\Delta\otimes id)\circ\Delta  $$
comme applications lin\'eaires de $V$ dans $V\otimes V \otimes V .$

La counit\'e  v\'erifiant:
$$(\epsilon\otimes id)\circ\Delta = (id\otimes\epsilon)\circ\Delta = id
:V\rightarrow V \ .$$

Etant donn\'ee une cog\`ebre : $ V(\Delta , \epsilon )$ ,  pour tout $v \in
V $ on notera:
$$\Delta (v) =\sum _{k} v_{k}^{'}\otimes  v_{k}^{''} \  .$$

\end{defin}

\begin{defin}[Big\`ebre .]\hspace{1cm}\\
\label{def1}
Une big\`ebre $A(\Delta , \epsilon )$ est une alg\`ebre associative  $A$ ,
avec unit\'e, et
munie d'une structure de cog\`ebre telle que le coproduit et la counit\'e
soient
des morphismes d'alg\`ebres .

\end{defin}

\begin{defin}[Alg\`ebre de Hopf .]\hspace{1cm}\\
\label{def1}
Une alg\`ebre de Hopf $H$  est une big\`ebre $H(\Delta , \epsilon ) $ munie
d'une
application lin\'eaire  $S:  H \rightarrow H$ qui v\'erifie pour tout $h$
dans $H$ :
$$\sum_{k} h_{k}^{'}.S( h_{k}^{''})= \epsilon (h) u_{H}$$
et
$$\sum_{k}S( h_{k}^{'}). h_{k}^{''}= \epsilon (h) u_{H}$$
o\`u   $u_{H}$ est l'unit\'e dans l'alg\`ebre $H$ et :
$$ \Delta (h) = \sum_{k} h_{k}^{'} \otimes  h_{k}^{''} \ . $$

\end{defin}
$S$ est appel\'ee l'antipode de l'alg\`ebre de Hopf . Elle est unique, et est
n\'ecessairement un anti-homomorphisme d'alg\`ebre et un anti-homomorphisme
de cog\`ebre.\\
\vspace{0.2cm}

Une r\'ef\'erence sur les structures d'alg\`ebres associatives, de cog\`ebres, 
big\`ebres, et  d'alg\`ebres de Hopf est [1];  pour les groupes
quantiques, et l' \'etude de la structure 
d'alg\`ebre de Hopf
quasi-triangulaire mise en \'evidence dans [4], pour l'\'etude de nombreux exemples, ainsi  
que pour avoir des r\'ef\'erences 
plus compl\`etes une r\'ef\'erence est: [2].\\
Le lemme 2.1 a \'et\'e publi\'e ant\'erieurement sous une forme
diff\'erente [3];
mais nous montrons dans cet article comment l'on peut l'utiliser pour
construire des
structures alg\'ebriques particuli\`eres.
\vspace{0.2cm}\\
Signalons encore que les big\`ebres obtenues ne sont pas n\'ecessairement
des alg\`ebres
finiment engendr\'ees .


\subsection{Big\`ebre libre construite sur une cog\`ebre .}
L'exemple typique et int\'eressant  de cog\`ebres, utiles pour la
construction de big\`ebres
pr\'esent\'ee dans cette note, est celui des cog\`ebres de dimensions finies
que nous
pr\'esentons pour \'eclairer les notations, mais il ne sera pas n\'ecessaire
de s'y
restreindre.\\    Soit
$ E$ une alg\`ebre associative sur
$C$ avec unit\'e, et de dimension finie; soit $F$ son dual ;\\
 $F$ est muni d'une structure de cog\`ebre. Le coproduit et la counit\'e sont
d\'efinis par:
$$\Delta : F \rightarrow F \otimes F \hspace{1em} et\hspace{1em}  \epsilon : F
\rightarrow C $$

$$\Delta f(e_1 \otimes e_2 )  = f(e_1.e_2) \hspace{1em} \forall f\in
F,\hspace{1em} et
\hspace{1em}
\forall e_1 , e_2
\in E$$

$$\epsilon (f) = f(u_{E}) \hspace{1em}\hbox{o\`u }
\hspace{1em}u_{E}\hspace{0.5em}est\hspace{0.5em} \hbox{ l'unit\'e}\hspace{0.5em}
dans\hspace{0.5em}  E \ .$$

La coassociativit\'e de $\Delta $ est assur\'ee par l'associativit\'e du
produit dans $E$:
$$(id\otimes\Delta)\circ \Delta = (\Delta\otimes id)\circ\Delta  $$
et
$$(\epsilon\otimes id)\circ\Delta = (id\otimes\epsilon)\circ\Delta = id
:F\rightarrow F \ .$$
Pour toute base $(e^{i})_{i\in I}$ de l'alg\`ebre $E$ en d\'esignant par
$(f_{i})_{i\in I}$  la base canonique duale de la cog\`ebre $F$ on a:
$$\Delta f_{i} =  \sum_{l,m} e_{i}^{l,m}f_{l}\otimes f_{m}$$
o\`u:
$$ e^{l}.e^{m} =\sum_{i} e_{i}^{l,m}e^{i}$$
et de plus l'on a :
$$u_{E} = \sum_{i} \epsilon (f_{i})e^{i} \ .$$

Dans ce qui suit nous ne supposerons plus que les cog\`ebres utilis\'ees
comme points de
d\'epart de la construction sont de dimensions finies.\\

Nous aurons besoin pour \'enoncer les propositions et les th\'eor\`emes par la suite ,
de d\'efinir  diff\'erents types de cog\`ebres .\\

\begin{defin}[Espace vectoriel \`a base d\'enombrable .]\hspace{1cm} 
C'est un espace vectoriel $F$ qui est la somme directe d'une famille d\'enombrable 
d'espaces vectoriels de dimensions finies $(F_{\alpha})_{\alpha \in N}$ .
\end{defin}
\begin{defin}[Formes \`a support fini sur un espace \`a base d\'enombrable.]
Une forme $\omega \in F^{*}$ est \`a support fini si $ \omega (F_{\alpha}) =0 $
sauf pour un nombre fini d'indices $\alpha $ .
\end{defin}

\begin{defin}[Cog\`ebre de type fini .]\hspace{1cm} 

Une cog\`ebre de type fini $F$ est un espace vectoriel \`a base d\'enombrable 
muni d'un coproduit coassociatif, et d'une counit\'e , telle que tout sous espace vectoriel
de dimension finie de $F$ soit contenu dans une sous cog\`ebre de dimension finie de $F$ .

\end{defin}
\vspace{0.2cm}
\begin{defin}[Cog\`ebre cofinie.]\hspace{1cm}
Une cog\`ebre cofinie est un espace vectoriel \`a base d\'enombrable muni d'un coproduit
coassociatif et d'une counit\'e , telle que pour tout $ f\in F$  \\ $\Delta (f) = \sum_{k} 
f_{k}^{'} \otimes f_{k}^{''}$ n'invoque q'une sommation finie , de termes lin\'eairement
ind\'ependants dans
$F\otimes F$ .

\end{defin}
\vspace{0.2cm}
\begin{defin}[Cog\`ebre r\'eguli\`ere .]\hspace{1cm}
Une cog\`ebre r\'eguli\`ere  est un espace vectoriel \`a base d\'enombrable 
muni d'un coproduit coassociatif et d'une counit\'e qui v\'erifient,\\
en notant pour tout $f\in F$ $$ \Delta (f) = \sum_{k} f_{k}^{'}\otimes f_{k}^{''}$$ 
et en supposant que les vecteurs $ f_{k}^{'} \otimes f_{k}^{''}$ sont lin\'eairement 
ind\'ependants dans $F\otimes F$ : \\
a) pour tout  $f \in  F $, et tout  sous espace vectoriel $F_{n} \subset F $ de dimension
finie,  l'ensemble des indices $k$, tels que
$0 \neq f_{k}^{'}\otimes f_{k}^{''} \in  F_{n}\otimes F $ ou $F\otimes F_{n}$ ,  est
fini.\vspace{0.2cm}\\ b) pour tout sous espace vectoriel $F_{n}$ de dimension finie,
l'ensemble des vecteurs
$f$ tels que $\forall k \  , \  C.f_{k}^{'}\otimes f_{k}^{''} \cap F_{n}\otimes F_{n}
={0}$ est un sous espace vectoriel de codimension finie .\\ c) pour tout $f\in F $
l'ensemble des indices
$k$ tels que 
$\epsilon
\otimes id(f_{k}^{'}\otimes f_{k}^{''})$ ou $ id \otimes
\epsilon(f_{k}^{'}\otimes f_{k}^{''})$ soient non nuls, est un ensemble fini .

\end{defin}

\vspace{0.3cm}

\begin{prop}[Big\`ebre libre construite sur une cog\`ebre $F$ . ]  
\hspace{1cm}

Soit $ F(\Delta,\epsilon) $ une cog\`ebre r\'eguli\`ere.\\
Soit $T(F)$ l'alg\`ebre
tensorielle construite sur l'espace vectoriel
$F$;
$$T(F)=C\oplus F\oplus F\otimes F\oplus\ldots\oplus\otimes^{n}F \oplus\ldots$$
Alors il existe un unique morphisme d'alg\`ebres, coassociatif:
$$\Delta : T(F) \rightarrow T(F)\otimes T(F)$$
et une unique counit\'e , qui soit un morphisme d'alg\`ebres :
$$\epsilon : T(F) \rightarrow C$$
qui v\'erifient sur $C \oplus F \subset T(F)$ :
 $$1\in C ; \hspace{0.5em}\Delta (1) = 1\otimes 1 ,\hspace{0.5em}\epsilon
(1) =1$$
$ \forall f\in F $,  $ \Delta (f) $ \hspace{0.3em} et\hspace{0.3em}
$\epsilon(f)$  co\"{\i}ncident avec
le coproduit et la counit\'e de la cog\`ebre $ F $ .
\end{prop}
D\'emonstration:
\\

Puisque $T(F)$ est l'alg\`ebre associative libre construite sur $F$
il n'y a pas d'obstructions \`a \'etendre par morphismes d'alg\`ebres
associatives
les applications lin\'eaires  $ \Delta $ et $\epsilon  $ d\'efinies de  $ F$
dans  $T(F)\otimes T(F)$ \hspace{0.4em} ou \hspace{0.4em} $C$; de plus les
relations
associ\'ees \`a l'identification de $1\in C$ \`a l'unit\'e de $T(F)$ sont
compatibles
avec $\Delta$ et $\epsilon$ .Finalement l'on a :
$$(id\otimes\Delta)\circ \Delta = (\Delta\otimes id)\circ\Delta  $$
et
$$(\epsilon\otimes id)\circ\Delta = (id\otimes\epsilon)\circ\Delta = id
:T(F)\rightarrow T(F) $$
parce qu'elles correspondent \`a des identit\'es entre morphismes d'alg\`ebres
qui co\"{\i}ncident sur les g\'en\'erateurs.
Ainsi l'image par le coproduit s'explicite:
$$ \Delta (f_1\otimes f_2\otimes\ldots\otimes f_n) =
\sum_{k1,k2,...,kn}(f_{k1}'\otimes f_{k2}'\otimes ..\otimes f_{kn}')\otimes
(f_{k1}''\otimes  f_{k2}''\otimes ..\otimes f_{kn}'')
$$
o\`u
$$\Delta (f_i) = \sum_{ki} f_{ki}^{'} \otimes f_{ki}^{''} .$$
On remarque que la big\`ebre tensorielle construite sur une cog\`ebre r\'eguli\`ere est 
encore une cog\`ebre r\'eguli\`ere ; elle est de plus la somme directe des sous cog\`ebres
 $\otimes ^{n}F$ , $n \geq 0$ .


\begin{rem}
Lorsque $F$ est la cog\`ebre duale d'une alg\`ebre $E$ de dimension finie,
le coproduit $$
\Delta : F\otimes F\otimes
\ldots
\otimes F
\rightarrow ( F\otimes F\otimes \ldots \otimes F)\otimes ( F\otimes
F\otimes \ldots \otimes
F)$$ correspond au coproduit  de la cog\`ebre duale de l'alg\`ebre
$$E\otimes E\otimes \ldots \otimes E$$
munie du produit ordinaire d\'efini sur un produit tensoriel d'alg\`ebres
.On a donc la
proposition suivante .
\end{rem}
\vspace{0.2cm}
\begin{prop}
Soit $E $ une alg\`ebre associative avec unit\'e, de dimension finie et $F$ la
cog\`ebre duale : $F =E^{*}$. La structure de cog\`ebre de  $T(F)$
d\'efinie par sa structure de big\`ebre  correspond \`a  celle d'une sous-cog\`ebre du
dual de l'alg\`ebre
$A(E)$,  produit directe  des alg\`ebres $\otimes ^{n} E $:
$$A(E)= \prod _{n\in N}\otimes ^{n} E \ .$$
\end{prop}
Remarquons simplement que pour tout n le sous espace  $T_{n}(F)$ suivant
est une sous
cog\`ebre de $T(F)$ :
$$ T_{n}(F) = C\oplus F\oplus \ldots \oplus \otimes^{n} F$$
et que l'alg\`ebre duale est  $A^{n}(E)$:
$$A^{n}(E) = C\oplus E \oplus \ldots \oplus  \otimes^{n}E $$
 son unit\'e \'etant:  $ 1_{C}+u_{E}+u_{E}\otimes u_{E}+\ldots
+\otimes^{n}u_{E}$ .

\subsection {Alg\`ebre des op\'erateurs invariants \`a droite sur une
cog\`ebre .}

\begin{defin}

Soit $F$ une cog\`ebre munie du coproduit $\Delta $ et de la counit\'e
$\epsilon $;
un op\'erateur lin\'eaire $$X: F\rightarrow F $$  est dit invariant \`a
droite s'
il v\'erifie:$$\Delta \circ X = (X\otimes id)\circ \Delta \  . $$ 
Ils d\'eterminent la sous alg\`ebre des op\'erateurs invariants \`a droite :
$$Inv_{d}(F,F) \subset Hom(F,F) \ .$$

\end{defin}
\vspace{0.5cm}
\begin{prop}
Soit $F$ une cog\`ebre cofinie ; 
alors l' ensemble des op\'erateurs invariants \`a droite  $ Inv_{d}(F,F) $, est une
sous alg\`ebre de l'alg\`ebre  $Hom(F,F)$ des op\'erateurs lin\'eaires sur $F$,
qui est anti-isomorphe en tant qu' alg\`ebre au dual $F^{*}$ de la cog\`ebre
coassociative  $F$. La bijection est donn\'ee par:
$$  i : X\in Inv_{d}(F,F) \rightarrow  i(X) =\epsilon \circ X \in F^{*} ,$$ 
l'application r\'eciproque par:$$i^{-1} : \omega \in F^{*} \rightarrow (\omega
\otimes id )\circ \Delta \hspace{0.3em} \in Inv_{d}(F,F) \hspace{0.3cm}.$$
Et la forme bilin\'eaire entre une cog\`ebre $ F(\Delta ,\epsilon )$ et
l'espace
$Inv_{d}(F,F)$ des op\'erateurs invariants \`a droite est donn\'ee par :
$$ <X,f>= \epsilon \circ X (f) .$$

\end{prop}
\vspace{0.3cm}
On remarque que dans le cas d'une cog\`ebre cofinie, l'identication des op\'erateurs
invariants \`a droite avec l'alg\`ebre (oppos\'ee) de $ F^{*} $ ne n\'ecessite pas
de pr\'ecautions sur le choix des formes dans $ F^{*} $ ; ce n'est d\'eja plus le
cas pour les cog\`ebres r\'eguli\`eres, o\`u dans cette note on se restreindra \`a un type
de formes tr\'es particulier .

\begin{prop}
Soit une cog\`ebre r\'eguli\`ere $F(\Delta ,\epsilon )$; c'est en particulier un espace
vectoriel \`a base d\'enombrable .\\ Alors l'espace vectoriel des formes \`a support fini
est muni d'une structure d'alg\`ebre associative anti-isomorphe \`a l'alg\`ebre des
op\'erateurs invariants \`a droite qu'elles  d\'efinissent :$Inv_{d,f}(F,F)$ .
Pour tout forme $\omega $ \`a support fini on a:
$$ X_{\omega}(f) = \omega \otimes id (\Delta (f) $$  et
$$ \omega (f) = \epsilon \circ X_{\omega} (f) . $$
\end{prop}
\vspace{0.3cm}
\begin{defin}[Formes r\'eguli\`eres. Op\'erateurs invariants \`a droite r\'eguliers. ]
Les op\'erateurs invariants \`a droite obtenus \`a partir de formes \`a support fini, et
l'op\'erateur identit\'e, sur une cog\`ebre r\'eguli\`ere $F$ seront dans la suite
d\'esign\'es sous le terme d'op\'erateurs invariants \`a droite r\'eguliers; l'alg\`ebre
sera d\'esign\'ee par :
$Inv_{d,r}(F,F)$.
\end{defin}
\vspace{0.3cm}
\begin{prop}[Groupe des \'el\'ements inversibles dans $Inv_{d,r}(F,F)$ .]
Soit $F(\Delta , \epsilon )$ une cog\`ebre r\'eguli\`ere et soit $B$ une sous alg\`ebre de
l'alg\`ebre des formes \`a support fini sur $F$;\\ 
supposont $B$ de dimension finie et avec unit\'e : $\epsilon _{B}$ .\\ Si $u \in B$  admet
$u^{-1}$  pour inverse dans $ B$ ,\\ alors $u + (\epsilon - \epsilon _{B})
$ est inversible dans l'alg\`ebre des formes r\'eguli\`eres et son inverse est :\\ $u^{-1} +
(\epsilon -
\epsilon _{B})$ .\\ L'ensemble des \'el\'ements inversibles obtenus de cette mani\`ere 
dans $Inv_{d,r}(F,F)$ engendre un groupe. 

\end{prop}

\vspace{0.2cm}
Dans la section 3 on aura  besoin d'utiliser les propri\'et\'es associ\'ees \`a des
cog\`ebres plus courantes .

\begin{defin}[Cog\`ebre fortement r\'eguli\`ere.]
C'est une cog\`ebre r\'eguli\`ere $F$ qui v\'erifie la propri\'et\'e suivante:\\
pour tout sous espace vectoriel $\Omega$ de dimension finie de formes $\in F^{*}$,  
\`a support fini, il existe un coid\'eal $J \subset F$, de codimension finie, tel que 
$$\Omega (J) =0$$
\end{defin}
\vspace{0.2cm}
\begin{prop}\hspace{0.3cm}Soit $F$ une cog\`ebre fortement r\'eguli\`ere et $X \subset
 Inv_{d,r}(F,F)$, un sous espace vectoriel de dimension fini d'op\'erateurs invariants \`a 
droite r\'eguliers agissant sur $F$, alors l'alg\`ebre engendr\'ee par $X$ est de
dimension finie .

\end{prop}

\vspace{0.3cm}
Le probl\`eme de la recherche d'une notion de formes plus g\'en\'erale sur une cog\`ebre
r\'eguli\`ere, ou fortement r\'eguli\`ere, et adapt\'ee au contexte de cette note n'est pas
abord\'e.\\

\vspace{0.3cm}
\begin{defin}

 Soit $E$ une alg\`ebre, son dual $E^{*}$ n'admet pas, en
g\'en\'eral, une structure de cog\`ebre mais on peut d\'efinir la notion de sous
cog\`ebre de
$E^{*}$: c'est une cog\`ebre $ F(\Delta ,\epsilon ) $ qui v\'erifie:
 $$ F(\Delta ,\epsilon )\subset E^{*} ,$$
$$ \forall f\in  F, \hspace{0.5em} f(e_{1}.e_{2}) = \Delta (f)(e_{1}\otimes
e_{2}))\hspace{0.2cm} ,$$

$$ \epsilon (f) = f(u_{E}) \hspace{0.2cm} .$$

\end{defin}

Alors on a:
\begin{prop} Soient  $ E $ une alg\`ebre et $ F(\Delta ,\epsilon ) $ une sous
cog\`ebre de $ E^{*} $ ; pour tout $  e \in E $ la transposition $ e.^{t}$
de l'op\'erateur multiplication \`a gauche,\\
$ e. :E \rightarrow E$ , est bien d\'efinie comme op\'erateur de $ F\rightarrow
F$,
et c'est un op\'erateur invariant \`a droite de $ F\rightarrow F $ et l'on a :
$$ i(e.^{t})(f) = f(e)  . $$

\end{prop}

\section{R\'ealisation explicite de big\`ebres .}
\vspace{0.3em}
Le lemme suivant donne un proc\'ed\'e pour construire des familles $(X_{i})_{i}$
 d'op\'erateurs lin\'eaires d'une alg\`ebre tensorielle $T(V)$  dans une
alg\`ebre
associative $B$, qui ont la propri\'et\'e suivante :

pour tout id\'eal $I(R)$ de $T(V)$ engendr\'e par un ensemble d'\'el\'ements
$R \subset T(V)$ alors : $$  X_{i}(R) = 0 \in B , \hspace{0.5em} \forall i
\hspace{0.3em}
\Rightarrow  X_{i}(I(R)) = 0 \in B , \hspace{0.5em}\forall i  . $$

\begin{lem}
Soit $V$ un espace vectoriel (somme directe d'une famille d\'enombrable
d'espaces vectoriels
de dimensions finies  ) , 
$T(V)$ l'alg\`ebre tensorielle associ\'ee, et
$B$ une alg\`ebre associative  avec unit\'e $1_{B}$.\\
Soit $L $ une cog\`ebre cofinie, de coproduit $\Delta $ , de counit\'e  $\epsilon $
, et $ x
$ une application lin\'eaire de  $L$ dans $Hom(V,B)$, l'espace des applications
lin\'eaires de $V$ dans $B$ :
$$ x : L \rightarrow Hom(V,B) $$

alors il existe une unique application lin\'eaire :
$$ X : L \rightarrow Hom(T(V), B )$$
qui v\'erifie:\\

1) pour $1 \in T(V)$,\hspace{0.3em}  $X(l)(1)= \epsilon (l)1_{B} $\\

2) pour $v\in V \subset T(V) $,\hspace{0.3em}  $X(l)(v) = x(l)(v) $ \\

3) pour tout couple  $ w_{1},w_{2}$  d'\'el\'ements dans $T(V)$:

$$X(l)(w_{1} \cdot w_{2}) = \sum_{k}
X(l_{k}^{'})(w_{1})._{B}X(l_{k}^{''})(w_{2})  $$
o\`u :
$$\Delta(l) = \sum_{k}l_{k}^{'}\otimes l_{k}^{''}\hspace{0.2cm}. $$

\end{lem}
D\'emonstration:\\

  Si $L$ n'est pas de dimension finie il est naturel de supposer que pour
tout $l \in  L$ ,  $\Delta (l) = \sum_{k} l_{k}^{'}\otimes l_{k}^{''} $ ,
n'invoque qu'une
sommation finie de termes lin\'eairement ind\'ependants dans $L\otimes L$ .

Notons par $\Delta ^{(n)} :  L \rightarrow
\otimes ^{n+1} L $ , l' application lin\'eaire 
obtenue par:
$$\otimes ^{n-1}(id)\otimes \Delta \circ \otimes ^{n-2}(id)\otimes \Delta
\circ \ldots \circ
\Delta  .$$
Alors  $ \otimes ^{n+1}(x) \circ \Delta ^{(n)} $ est une application lin\'eaire de\\
$L $ dans
$Hom( \otimes ^{n+1}V \rightarrow \otimes ^{n+1} B )$ ; finalement en
composant avec le
produit ordonn\'e : $\otimes ^{n+1}B \rightarrow B $,  pour chaque  $l$ , on
obtient un
op\'erateur
 not\'e $X_{n+1}(l)$, dans $Hom( \otimes ^{n+1}V ,B )$,  et par lin\'earit\'e
on d\'efinit l'op\'erateur :
 $$ X(l) :  V\oplus V\otimes V\oplus\ldots\oplus\otimes^{n}V
\oplus \ldots \rightarrow B .  $$

Ils v\'erifient pour $w_{1} \in \otimes ^{p}V $ et, $w_{2} \in \otimes ^{q}V $,
$ w_{1}\cdot w_{2}  \in \otimes ^{p+q}V $ :

$$ X(l) (w_{1}\cdot w_{2}) = \sum_{k} X(l_{k}^{'})(w_{1}).X(l_{k}^{''})(w_{2})
\hspace{0.3cm}.
$$

Cette propri\'et\'e est la  cons\'equence de la coassociativit\'e du
coproduit sur $L$ .

De plus on la pr\'eserve sur $T(L)$ en imposant sur  $C $ : $ X(l)(1)
=\hspace{0.3em}
\epsilon (l).1_{B}$  .

\begin{thm}
Soit $ L(\Delta _{L},\epsilon _{L}) $ une cog\`ebre cofinie,  $F(\Delta _{F},\epsilon
_{F})$ une cog\`ebre r\'eguli\`ere, 
et $x$ une application lin\'eaire de la cog\`ebre $L$ dans l'alg\`ebre des
op\'erateurs
invariants
\`a droite agissant sur la cog\`ebre $F$ . On consid\`ere $F$ comme \'etant
inclus dans
l'alg\`ebre associative avec unit\'e
$T(F)$.
 $$x :L \rightarrow Inv_{d}(F,F ) $$
d\'efinit une application lin\'eaire de $L$ dans $Hom(F,T(F))$, et  le
lemme 2.1 d\'efinit donc pour chaque $l\in L$ un op\'erateur $X(l)$  :
$$ X(l) : T(F) \rightarrow T(F) \hspace{0.4em} . $$

Ils v\'erifient :\\

1) $ X(l)(1_{T(F)}) =\epsilon _{L}(l).1_{T(F)} $ ;\\

2)   $X(l)(f) = x(l)(f)  \hspace{0.5em} \forall f\in F $ ;\\

3)  Pour tout $w_{1},w_{2}\hspace{0.3cm}\in T(F) $ 
 $$  X(l) (w_{1}\cdot w_{2}) = \sum_{k}
X(l_{k}^{'})(w_{1})\cdot X(l_{k}^{''})(w_{2})$$
o\`u $\Delta_{L}(l) =\sum_{k}l_{k}^{'}\otimes l_{k}^{''}$ ;\\

4)  Pour tout $n \geq 0$, \hspace{0.3cm} $X(l) :
\otimes ^{n}F \rightarrow \otimes ^{n}F $ ;\\

5) Les op\'erateurs $X(l)$ sont des op\'erateurs
invariants \`a droite sur la big\`ebre
 libre   $T(F)$ \hspace{0.2cm}.

\end{thm}
D\'emonstration:\\

Les propri\'et\'es 1,2,3 proviennent du lemme 2.1 et la propri\'et\'e 4  est une
cons\'equence du fait que : $x(l) :F\rightarrow F $ transforme  simplement
$F$ dans $F$
et des propri\'et\'es 1 et 3 .
D\'emontrons la propri\'et\'e 5  : \\

Soit la big\`ebre libre $T(F)$, notons par $\Delta _{F}$ et $\epsilon _{F}$
son coproduit
et sa counit\'e, qui sont des morphismes d'alg\`ebres respectivement de \\
$T(F)\rightarrow  T(F)\otimes T(F)$ et de  $T(F)\rightarrow C$ ; notons
encore par $\Delta
_{L}$ le coproduit sur la cog\`ebre  $L$ et $\epsilon _{L}$ la counit\'e;
soit :$$ Y(l)
=\Delta_{F}
\circ X(l)
\hspace{0.5em},
\hspace{0.5em}Z(l) = (X(l)\otimes id)\circ \Delta _{F} \ .$$

Ce sont des op\'erateurs de $T(F)$ dans $T(F)\otimes T(F)$  qui v\'erifient:
$$Y(l)(1_{T(F)}) =\epsilon_{L} (l).1_{T(F)}\otimes1_{T(F)}= Z(l)(1_{T(F)}) \  ;$$
par hypoth\`ese les op\'erateurs $x(l)$ sont des op\'erateurs invariants
\`a droite sur $F$

et donc :
$$ \forall f \in F ,\hspace{0.3cm} Y(l)(f) =Z(l)(f) \ .$$
D' autre part pour tout $ w_{1},w_{2} \in T(F)$ on a :
$$ Y(l)(w_{1}\cdot w_{2}) =\sum_{k}Y(l_{k}^{'})(w_{1}) ._{T(F)\otimes T(F)}
Y(l_{k}^{''})(w_{2})        $$
 les op\'erateurs $Z(l)$ v\'erifient aussi:
$$ Z(l)(w_{1}\cdot w_{2}) =\sum_{k}Z(l_{k}^{'})(w_{1}) ._{T(F)\otimes T(F)}
Z(l_{k}^{''})(w_{2})        $$
o\`u :

$$\sum_{k}l_{k}^{'}\otimes l_{k}^{''} =\Delta _{L} (l) \ .$$
Les op\'erateurs  $X(l)$ sont donc bien des op\'erateurs invariants \`a
droite sur la
big\`ebre $T(F)(\Delta _{F},\epsilon _{F})$, parce que les op\'erateurs
$Y(l)$ et $Z(l)$
co\"{\i}ncident .\\

Il est important pour avoir un coproduit sur la sous alg\`ebre engendr\'ee
par l'op\'erateur
identit\'e et la famille d' op\'erateurs $(X(l))_{l \in L}\hspace{0.5em} \subset
Hom(T(F),T(F))$, que ces op\'erateurs soient des op\'erateurs invariants
\`a droite sur la
big\`ebre $T(F)(\Delta _{F},\epsilon _{F})$ .Ceci est d\'emontr\'e dans le
th\'eor\`eme
suivant sous une condition plus restrictive: $x(L) \subset Inv_{d,r}(F,F)$.

\begin{thm}
Soit $L$ et $F$ deux cog\`ebres v\'erifiant   :\\
a) $L$ est une cog\`ebre cofinie ;\\
b) $F$ est une cog\`ebre r\'eguli\`ere ; 
\\

Soit  $x :L \rightarrow Inv_{d,r}(F,F) $ une application lin\'eaire de la
cog\`ebre $L$
dans les op\'erateurs invariants \`a droite r\'eguliers agissant sur la cog\`ebre $F$ et
soit :\\

$$X(l) \in Inv_{d}(T(F),T(F) \subset Hom(T(F),T(F))$$
les op\'erateurs donn\'es par le th\'eor\`eme 2.1\\

Alors l'alg\`ebre $U_{x}$, engendr\'ee par les op\'erateurs $X(l)$, $l\in L$ et
l'identit\'e 
est munie d'une unique structure de big\`ebre $U_{x}(\Delta _{L},\epsilon
_{L})$ telle que
:\\

$\Delta _{L} :U_{x} \rightarrow U_{x} \otimes U_{x}$ \'etend par morphisme
d'alg\`ebres
le coproduit d\'efini sur les g\'en\'erateurs par :
$$ \Delta _{L} : X(l) \rightarrow \sum_{k}X(l_{k}^{'})\otimes X(l_{k}^{''})$$
$$\Delta _{L} : 1_{U_{x}} \rightarrow 1_{U_{x}}\otimes 1_{U_{x}}$$
o\`u  $$\sum_{k}l_{k}^{'}\otimes l_{k}^{''} =\Delta _{L}(l)$$
et la counit\'e $\epsilon _{L} :U_{x} \rightarrow C$ \'etend par morphisme
d'alg\`ebres
la counit\'e d\'efinie sur les g\'en\'erateurs par:$$\epsilon _{L}
(1_{U_{x}})=1
,\hspace{0.3em}
\epsilon _{L} (X(l))= \epsilon _{L}(l) .$$

\end{thm}

D\'emonstration:

pour un mon\^ome form\'e avec les g\'en\'erateurs de $U_{x}$ nous avons, pour tout 
couple $w_{1} , w_{2} \in T(F) $ :
 $$ X(l_{1})\circ X(l_{2})\circ \ldots \circ X(l_{n}) (w_{1}\cdot w_{2}) $$
est \'egal \`a :

$$\sum_{k1,k2,..,kn} X(l_{k1}^{'}) \circ  X(l_{k2}^{'}) \circ  \ldots
 \circ X(l_{kn}^{'})(w_{1}) \cdot   X(l_{k1}^{''}) \circ  X(l_{k2}^{''})
\circ \ldots
 \circ X(l_{kn}^{''})(w_{2}) \ .$$

Ainsi \`a tout polyn\^ome $Z$ form\'e sur les g\'en\'erateurs $1$ et
$X(l))$ on associe par
lin\'earit\'e un \'el\'ement de $U_{x}\otimes U_{x} $ qui satisfait la
propri\'et\'e
pr\'ec\'edente; le point est qu'en g\'en\'eral ceci ne d\'efinit pas une
application de \\
$U_{x}
\rightarrow U_{x}\otimes U_{x}$ , puisque un m\^eme op\'erateur est
repr\'esent\'e  par des
polyn\^omes diff\'erents ; d\'emontrons dans notre cas que l'on a bien une
application de
$U_{x}
\rightarrow U_{x}\otimes U_{x}$ qui sera alors clairement un morphisme
d'alg\`ebres.
Supposons\\ $Z (w) = 0 $ ,  $\hspace{0.3em} \forall \hspace{0.3cm} w \in T(F))$; Soit un
\'el\'ement
 $\sum_{k}Z_{k}^{'}\otimes Z_{k}^{''} \in U_{x}\otimes U_{x}$  tel que :
$$ Z(w_{1}\cdot w_{2})= \sum_{k} Z_{k}^{'}(w_{1})\cdot  Z_{k}^{''}(w_{2})
,\hspace{0.3cm}
\forall w_{1},w_{2} \in T(F)  . $$
Montrons que pour tout \'el\'ement $w_{1} \otimes w_{2} \in T(F)\otimes T(F)$
$$\sum_{k }Z_{k}^{'}(w_{1}) \otimes  Z_{k}^{''}(w_{2}) = 0 \hspace{0.4cm}
dans\hspace{0.3cm}
T(F)\otimes T(F) .$$
F \'etant une cog\`ebre r\'eguli\`ere $T(F)$ est une big\`ebre (r\'eguli\`ere) et nous
avons:$$
\forall w\in T(F) ,  w= (\epsilon
\otimes
id)\Delta (w)$$
Les op\'erateurs $Z_{k}^{'}$  et  $   Z_{k}^{''}$ \'etant des op\'erateurs
invariants \`a
droite r\'eguliers sur $T(F)$ il s'en d\'eduit que:
$$\sum_{k} Z_{k}^{'}(w_{1})\otimes    Z_{k}^{''}(w_{2})$$
est donn\'e par
 $$\sum_{k} (\epsilon \circ Z_{k}^{'}\otimes id ) \Delta (w_{1})\otimes
(\epsilon \circ
Z_{k}^{''}\otimes id ) \Delta (w_{2}). $$
D'autre part notons:
$$ \Delta (w_{1}) = \sum_{i} w_{1,i}^{'}\otimes w_{1,i}^{''}$$

$$ \Delta (w_{2}) = \sum_{j} w_{2,j}^{'}\otimes w_{2,j}^{''}$$
et on obtient :
$$\sum_{k,i,j} \epsilon \circ Z_{k}^{'} (w_{1,i}^{'}).w_{1,i}^{''}\otimes
 \epsilon \circ Z_{k}^{''}(w_{2,j}^{'}).w_{2,j}^{''} $$
o\`u les sommations sur les indices $i,j$ s'effectuent sur un nombre fini de termes
non nuls.
En effectuant d'abord la sommation en k et parce que $\epsilon $ est un
morphisme
$T(F) \rightarrow C $ on obtient $0$ comme cons\'equence  de l'hypoth\`ese\\
$Z:T(F) \rightarrow T(F) =0$ ;  en effet les sommes suivantes:

$$  \sum_{k} \epsilon \circ Z_{k}^{'} (w_{1,i}^{'}).\epsilon \circ
Z_{k}^{''}(w_{2,j}^{'})
$$
s'\'ecrivent:  $ \sum_{k}\epsilon( Z_{k}^{'} (w_{1,i}^{'}).
Z_{k}^{''}(w_{2,j}^{'})) =
\epsilon \circ Z(w_{1,i}^{'}.w_{2,j}^{'}) =0  .$
\\

Ainsi $\Delta _{L}$ est bien d\'efini et c'est clairement un morphisme
d'alg\`ebres ;\\

Pour ce qui concerne la counit\'e , on remarque que les op\'erateurs de
l'alg\`ebre
$U_{x}$
 laissent invariant le sous espace  $C \subset  T(F)$ ; et en d\'efinissant
pour $u \in
U_{x}$
$$\epsilon_{L}(u) .1= u(1)$$
on obtient un morphisme $ U_{x}\rightarrow C $  qui satisfait les
conditions sur les
g\'en\'erateurs $ X(l) , l\in L $ et   $1$ .
Les autres propri\'et\'es de la structure de big\`ebre en d\'ecoulent.\\

Le th\'eor\`eme suivant donne les conditions qui permettent d'obtenir une
information 
plus pr\'ecise sur l'id\'eal des relations des big\`ebres obtenues dans le th\'eor\`eme 
2.2  .

\begin{thm}
Soit $L$ et $F$ deux cog\`ebres v\'erifiant :\\
a) $L$ est une cog\`ebre de type fini.\\
b) $F$ est une cog\`ebre fortement r\'eguli\`ere.
\\

Soit  $x :L \rightarrow Inv_{d,r}(F,F) $  une application lin\'eaire de la
cog\`ebre $L$
dans les op\'erateurs invariants \`a droite r\'eguliers sur la cog\`ebre $F$  .\\

Soit $U_{x}(\Delta_{L},\epsilon_{L})$  la big\`ebre donn\'ee par le th\'eor\`eme 2.2
engendr\'ee par l'identit\'e de $T(F)$ dans $T(F)$ et les op\'erateurs $X(l) $ :

$$X(l) \in Inv_{d}(T(F),T(F) \subset Hom(T(F),T(F)) \ .$$

Alors:  \\
soit le morphisme d'alg\`ebres ,   
$ \pi : T(L) \rightarrow U_{x} $ d\'efini sur les mon\^omes par :
$$ \pi (l_{i1}^{j1}\otimes \ldots \otimes l_{in}^{jn})=  X(l_{i1}^{j1})\circ
 \ldots
\circ X(l_{in}^{jn}) \ ;$$
il d\'efinit une action et en particulier une repr\'esentation de
$T(L)(\Delta _{L},\epsilon _{L})$  sur
$T(F)$.\\

 L'id\'eal des relations $I_{x}= ker\pi  \subset T(L) $ est engendr\'e par une r\'eunion
d\'enombrable $(R_{\alpha})_{\alpha \in N}$ d'espaces vectoriels de dimensions finies qui
sont chacun des coid\'eaux  de la big\`ebre $T(L)(\Delta _{L};\epsilon _{L})$ .\\

\end{thm}
La d\'emonstration repose sur les hypoth\`eses , que $L$ soit une cog\`ebre de type fini,
et que $F$ soit une cog\`ebre fortement r\'eguli\`ere ; ceci ayant en particulier pour
cons\'equence (proposition 1.6) que les op\'erateurs $ X(L_{n}) \subset Inv_{d,r}(F,F) $
engendrent une alg\`ebre de dimension finie, lorsque $L_{n}$ est un sous espace vectoriel de
dimension finie de la cog\`ebre $L$ .

\section{R\'ealisation de big\`ebres pr\'esentant une alg\`ebre
de Hopf associ\'ee .}

Soit l'alg\`ebre $ K^{+} $ des matrices infinies vivant dans le
demi-secteur sup\'erieur.
Une matrice $ (m_{i}^{j})$ est dans $ K^{+} $ si  $
m_{i}^{j}=0 $ lorsque $j $ est strictement sup\'erieur \`a $i$.  Soit $
L^{+}$  la cog\`ebre
duale;  une base est form\'ee par les
\'el\'ements de la suite
$l_{i}^{j}$ avec $ i,j \in N $  et $j\leq i$ ; le coproduit et la
co-unit\'e sont d\'efinis
par :
$$\Delta (l_{i}^{j}) =\sum_{k \in (j,i)} l_{k}^{j} \otimes l_{i}^{k} $$
$$ \epsilon (l_{i}^{j}) = \delta (j,i) $$

Soit d'autre part les cog\`ebres :
$$ L_{n}^{+} \hspace{0.2cm}\hbox{ d\'efinies\hspace{0.2cm} pour\hspace{0.2cm}
tout}\hspace{0.2cm}
n  \in \hspace{0.2cm}N  $$ comme \'etant les cog\`ebres duales des
alg\`ebres $ M_{n}^{+}
$  de matrices trigonales sup\'erieures.

\begin{defin}
On dira qu'une cog\`ebre $L$ est cotrigonale si c'est une somme directe de
cog\`ebres
d\'efinies ci-dessus. De plus on d\'esignera par $D_{L}$ l'ensemble obtenu
par la r\'eunion
des \'el\'ements  $l_{i}^{i}$ diagonaux de chaque cog\`ebres intervenant
dans la somme
directe.
\end{defin}
On a pour tout : $$l\in D_{L} : \hspace{0.3cm}\epsilon (l) =1 \hspace{0.3cm}et
\hspace{0.3cm}
\Delta (l)= l\otimes l$$
\vspace{1cm}
\begin{thm}Soit $F$ une cog\`ebre fortement r\'eguli\`ere
et $L$ une cog\`ebre  co-trigonale;\\
soit $x :L \rightarrow  Inv_{d,r}(F,F) $ une application lin\'eaire de $L$ dans
l'alg\`ebre des op\'erateurs invariants \`a droite r\'eguliers, v\'erifiant:\\

a)pour tout $l\in D_{L}$ ,  $x(l)$ est un op\'erateur inversible dans
$Inv_{d,r}(F,F)$ \\

b) pour tout $l\in D_{L} $  il existe $l^{'}\in D_{L}  $ tel que $$x(l)
\circ x(l^{'})= id
:F \rightarrow F .$$
Alors : \\

 A) Soit la big\`ebre $U_{x}(\Delta_{L},\epsilon _{L})\hspace{0.3cm}  \subset
\hspace{0.3cm} Inv_{d}(T(F),T(F))$, engendr\'ee par les op\'erateurs $X(l)$
 et l'identit\'e : \\

il existe une solution unique dans l'alg\`ebre $U_{x}$ ,
$Y_{i}^{j}, \hspace{0.3cm}j\leq  i$  au syst\`eme d'\'equations  :
$$  \sum_{k} X(l_{k}^{j})\circ  Y_{i}^{k} =\epsilon_{L} ( X(l_{i}^{j})).1_{U_{x}}
=\delta (i,j).1_{U_{x}}$$
et
$$ \sum_{k} Y_{k}^{j}\circ  X(l_{i}^{k}) =\epsilon_{L} ( X(l_{i}^{j})).1_{U_{x}}
=\delta (i,j) .1_{U_{x}} \hspace{0.2cm} .$$
De plus chaque

$$  Y_{i}^{j} $$
est donn\'e par un polyn\^ome explicite et fini dans les g\'en\'erateurs
$X(l_{l}^{m})$ .\\

B)  Soit le morphisme d'alg\`ebres ,   
$ \pi : T(L) \rightarrow U_{x} $ d\'efini par :
$$ \pi (l_{i1}^{j1}\otimes \ldots \otimes l_{in}^{jn})=  X(l_{i1}^{j1})\circ
 \ldots
\circ X(l_{in}^{jn}) \ ;$$
il d\'efinit une action et en particulier une repr\'esentation de $T(L)$  sur $T(F)$.
L'id\'eal des relations $I_{x} \subset T(L) $ est engendr\'e par une r\'eunion
d\'enombrable $R$ d'espaces vectoriels de dimensions finies qui sont chacun des
coid\'eaux  de la big\`ebre $T(L)(\Delta _{L};\epsilon _{L})$ .\\

C)  La condition n\'ecessaire et suffisante pour que l'antipode existe sur
la big\`ebre
 $U_{x}(\Delta_{L},\epsilon _{L})$  est que l'id\'eal bilat\`ere $I_{x}$ des
relations soit
engendr\'e par un ensemble $R$ de relations qui soit stable par
l'anti-homomorphisme alors
d\'efini  $ S :T(L) \rightarrow T(L) $ , v\'erifiant $\pi \circ S(l_{i}^{j}) = Y_{i}^{j}$ .
\end{thm}
D\'emonstration:\\

On se restreint au cas o\`u la cog\`ebre cotrigonale $L$ se r\'eduit \`a
la cog\`ebre
$L^{+}$ .

Soit la matrice $ M_{x}$ \`a coefficients dans $Inv_{d}(T(F),T(F))$
d\'efinie par:

$$m_{i}^{j}= X(l_{i}^{j}) \hspace{0.3cm} j\leq i ;$$ il est facile de
v\'erifier que cette
matrice est inversible, et qu' en particulier les coefficients  $Y_{i}^{j} \in
Inv_{d}(T(F),T(F))$  de la matrice inverse s'obtiennent comme polyn\^omes
finis explicites
dans les  op\'erateurs $  X(l_{l}^{m})$ pour
$j\leq i$ .Ceci repose sur les faits que les op\'erateurs sur la diagonale
$X(l_{i}^{i})$
sont tous inversibles et que la matrice $(X_{i}^{j})$ est de structure
triangulaire .

Donc nous obtenons l'unique solution (dans $U_{x}(\Delta_{L},\epsilon_{L})$ ) au
probl\`eme suivant :
$$ \sum_{k} X(l_{k}^{j})\circ  Y_{i}^{k} =\epsilon_{L} ( X(l_{i}^{j})). 1_{U_{x}} 
=\delta (i,j). 1_{U_{x}}$$
et
$$ \sum_{k} Y_{k}^{j}\circ  X(l_{i}^{k}) =\epsilon_{L} ( X(l_{i}^{j})). 1_{U_{x}}
=\delta (i,j) . 1_{U_{x}}$$

ce qui d\'efinit l'application $S_{1} (X(l_{i}^{j})  = Y_{i}^{j} $ .\\

L'id\'eal des relations de la big\`ebre $U_{x}(\Delta _{L};\epsilon _{L})$, se d\'ecrit de
la mani\`ere suivante:\\
l'application lin\'eaire $\pi: T(L) \rightarrow Inv_{d}(T(F),T(F))$ donn\'ee par
$$ \pi (l_{i1}^{j1}\otimes \ldots \otimes l_{in}^{jn})=  X(l_{i1}^{j1})\circ
 \ldots
\circ X(l_{in}^{jn})$$
est une action de $T(L) $ sur $T(F)$ et en particulier une repr\'esentation dont l'id\'eal
associ\'e  $I_{x}\subset T(L)$ caract\'erise les relations dans $U_{x}$ .\\

On montre parceque d'une part $L$ est une cog\`ebre cotrigonale et donc de type fini,
et d'autre part , $F$ est une cog\`ebre fortement r\'eguli\`ere, que l'id\'eal $I_{x}$
est engendr\'e par une r\'eunion d\'enombrable d'espaces vectoriels de dimensions finies
qui sont chacun des coid\'eaux de $T(L)(\Delta_{L},\epsilon _{L})$ .\\ 

La condition n\'ecessaire et
suffisante pour avoir une
antipode sur
$U_{x}(\Delta _{L},\epsilon _{L} )$  est que les relations soient
compatibles  avec
l'anti-homomorphisme a priori d\'efini sur les g\'en\'erateurs; ceci finit
la d\'emonstration
du th\'eor\`eme 3.1 .\\

Mais en fait on va d\'emontrer des propri\'et\'es plus fortes dans la
big\`ebre
$U_{x}(\Delta_{L},\epsilon _{L})$  qui conduisent au th\'eor\`eme 3.2  .
\\

Si l'on choisit une d\'etermination particuli\`ere de l'inverse de
l'op\'erateur diagonal
$X_{i}^{i} ,(i,i) \in D_{L} $ on obtient alors une application $S_{1}^{r} :
L \rightarrow
T(L)
$ qui v\'erifie :
$$ \pi ( S_{1}^{r}(l_{i}^{j}) =Y_{i}^{j} $$
o\`u $\pi$ est le morphisme d'alg\`ebres, de $T(L)$ dans  $U_{x} $  , d\'efini par :
$$ \pi (l_{i1}^{j1}\otimes \ldots \otimes l_{in}^{jn})=  X(l_{i1}^{j1})\circ
 \ldots
\circ X(l_{in}^{jn}) \ .$$
Soit $ S^{r}$ l'application lin\'eaire de $T(L)$ dans $T(L) $ obtenue en
prolongeant $
S_{1}^{r} $ par anti-homomorphisme .\\

D'autre part on montre que les $Y_{i}^{j}$ sont uniquement d\'etermin\'es
comme solution
du probl\`eme pr\'ec\'edent et qu'ils satisfont:
$$\Delta_{L} Y_{i}^{j}= \sum_{k\in (j,i)} Y_{i}^{k}   \otimes Y_{k}^{j} ; $$
\\
on montre de plus que dans l'espace vectoriel des mon\^omes d'ordre n \\

 $ w =l_{i1}^{j1}\otimes
\ldots
\otimes l_{in}^{jn}
\in
 \otimes ^{n}(L)   $ :\\

la solution du probl\`eme :

$$\sum_{k1,..kn} X(l_{k1}^{j1})\circ \ldots \circ X(l_{kn}^{jn}) \circ
Y_{i1,..,in}^{k1,..,kn}= \epsilon (  X(l_{i1}^{j1})\circ \ldots \circ
X(l_{in}^{jn})) . 1_{U_{x}}
$$
est unique dans la big\`ebre  $U_{x}$ et est donn\'ee par:
$$  Y_{i1,..,in}^{k1,..,kn}=Y_{in}^{kn}\circ ..\circ Y_{i1}^{k1} =\pi
(S^{r}(l_{i1}^{k1}\otimes \ldots  \otimes l_{in}^{kn}     )  \  .$$

\vspace{0.8cm}

\vspace{0.8cm}

\begin{thm}Soit $F$ une cog\`ebre fortement r\'eguli\`ere,  $L$ une cog\`ebre  co-trigonale
; soit $x :L \rightarrow  Inv_{d,r}(F,F) $ une application lin\'eaire de $L$ dans
$Inv_{d,r}(F,F)$ qui v\'erifie:\\

a)pour tout $l\in D_{L}$ ,  $x(l)$ est un op\'erateur inversible dans
$Inv_{d,r}(F,F)$ \\

b) pour tout $l\in D_{L} $  il existe $l^{'}\in D_{L}$ tel que $$x(l) \circ
x(l^{'})= id
:F \rightarrow F .$$
Alors : \\

Soit $T(L)(\Delta _{L} , \epsilon _{L})$ la big\`ebre universelle et soit $
I_{x}$
l'id\'eal associ\'e par le morphisme d'alg\`ebres $\pi :T(L) \rightarrow U_{x}$ d\'efini par
:
$$ \pi : l_{i1}^{j1}\otimes l_{i2}^{j2}\otimes \ldots \otimes l_{in}^{jn}\rightarrow
X(l_{i1}^{j1})\circ X(l_{i2}^{j2})\circ \ldots \circ X(l_{in}^{jn}) $$

Soit $ S^{r}$ de $T(L) \rightarrow T(L) $ l'application lin\'eaire
pr\'ec\'edemment
d\'efinie .\\

Alors :\\

A)  l'id\'eal $I_{x}$  est un coid\'eal;\\

B) il existe un id\'eal minimal  $J_{x} $  de la big\`ebre libre  $T(L)(\Delta
_{L},\epsilon _{L})
$
  v\'erifiant :\\

1) $I_{x}\subset  J_{x}$\\

2) $ J_{x}$ est laiss\'e invariant par $S^{r}$\\

3) $ J_{x}$ est un coid\'eal de la big\`ebre  $T(L)(\Delta _{L} ,\epsilon
_{L} )$ .\\

C)   $ H_{x} = T(L)(\Delta _{L},\epsilon _{L})/J_{x} $   est l' alg\`ebre
de Hopf,
canoniquement associ\'ee \`a la big\`ebre $U_{x}(\Delta _{L},\epsilon _{L}) $.

\end{thm}

D\'emonstration:\\
On admettra dans cette note  que l'id\'eal $I_{x}$ est aussi un coid\'eal :
$$\Delta _{L} I_{x} \subset I_{x} \otimes T(L) +T(L) \otimes I_{x} .$$
Dans le th\'eor\`eme 3.1 on a pr\'ecis\'e les hypoth\`eses qui permettent de le d\'eduire:
$L$ est une cog\`ebre de type fini , $F$ est une cog\`ebre fortement r\'eguli\`ere,
et $x $ une application lin\'eaire de $L$ dans les op\'erateurs invariants \`a droite
r\'eguliers sur $F$ .

D'autre part montrons que pour tout mon\^ome $ l_{i1}^{j1}\otimes \ldots \otimes
l_{in}^{jn}
$ l'on
a :
$$\Delta_{L} \circ S^{r}(l_{i1}^{j1}\otimes \ldots \otimes l_{in}^{jn} )
\in I_{x}\otimes
T(L) + T(L)\otimes I_{x} + S^{r}\otimes S^{r}\circ \Delta _{L}
^{op}(l_{i1}^{j1}\otimes
\ldots
\otimes l_{in}^{jn} ) \ .$$
En effet:
$$ \pi \circ S^{r}(l_{i1}^{j1}\otimes \ldots \otimes l_{in}^{jn}) =
Y_{in}^{jn}\circ ..
\circ Y_{i1}^{j1}$$
$$\Delta _{L} ( Y_{in}^{jn}\circ .. \circ Y_{i1}^{j1}) =
\sum_{k1,..,kn} Y_{in}^{kn}\circ ..\circ Y_{i1}^{k1}\otimes Y_{kn}^{jn}\circ
..\circ Y_{k1}^{j1} \ ;$$ ce qui est \'egal dans $U_{x}\otimes U_{x}$ \`a :
$$( \pi \otimes \pi)\circ ( S^{r}\otimes S^{r })\Delta
_{L}^{op}(l_{i1}^{j1} \otimes
..\otimes l_{in}^{jn}) \ ;
$$
donc pour tout \'el\'ement $ w \in T(L)$ nous avons :
$$\Delta _{L} (S^{r}( w))  =  S^{r} \otimes  S^{r}\circ \Delta
_{L}^{op}(w)\hspace{0.5cm} modulo \hspace{0.5cm}I_{x}\otimes T(L) +
T(L)\otimes I_{x} \ .
$$

Soit $R$ un sous-espace vectoriel de $T(L)$  qui v\'erifie:
$$\Delta _{L}  R \subset R \otimes T(L) +T(L) \otimes R .$$

Alors, $$ \Delta _{L} S^{r}(R) \subset  S^{r}(R) \otimes T(L) +T(L) \otimes
S^{r}(R) +
I_{x}\otimes T(L) + T(L)\otimes I_{x} \ . $$

Soit $R_{0}$  un sous espace vectoriel de relations qui d\'etermine $I_{x}$
(l'id\'eal
bilat\`ere engendr\'e par $R_{0}$ et l'id\'eal $I_{x}$ co\"{\i}ncident),  et  qui soit   un
coid\'eal de
la big\`ebre $T(L)(\Delta _{L} , \epsilon _{L})$ ;\\

Soit $ R_{n}= \sum _{i\in ( 0,n)} (S^{r}) ^{(i)}(R_{0})$ :
  par r\'ecurrence  on montre que  $R_{n}$  v\'erifie :
$$\Delta _{L} S^{r}(R_{n}) \subset  I(R_{n+1})\otimes T(L) +T(L)\otimes
I(R_{n+1})$$
et donc:
$$\Delta _{L} (R_{n+1}) \subset  I(R_{n+1})\otimes T(L) +T(L)\otimes
I(R_{n+1})$$
o\`u $I(R_{n}) $d\'esigne l'id\'eal bilat\`ere engendr\'e par $R_{n}$ .\\

$\cup _{n\in N} R_{n}$ est donc le plus petit sous espace vectoriel de
$T(L)$ stable par
 $S^{r}$ , contenant  $R_{0} $  ; et l'id\'eal engendr\'e est bien un
coid\'eal de la
big\`ebre : $ T(L)(\Delta _{L},\epsilon _{L})$   .\\

L'alg\`ebre $ H_{x} = T(L)(\Delta _{L},\epsilon _{L}) / J_{x}$ est une alg\`ebre de Hopf
canoniquement associ\'ee  \`a la big\`ebre $U_{x}$ du th\'eor\`eme 3.2  .\\

\section{Crit\`ere pour qu'une big\`ebre r\'ealis\'ee admette une alg\`ebre de Hopf
associ\'ee .}

Le th\'eor\`eme suivant, plus abstrait, est \'enonc\'e en partie pour
\'eviter toutes
confusions dans les \'etapes du th\'eor\`eme 3.2 d'une part , et  d'autre
part pour donner
une condition abstraite qui permet de s'affranchir de la condition
impos\'ee dans les
th\'eor\`emes 3.1 et 3.2 sur la cog\`ebre  $L$  d'\^etre cotrigonale .

\begin{thm}
Soit $L$ et $F$ deux cog\`ebres v\'erifiant :\\
a) $L$ est une cog\`ebre de type fini.\\
b) $F$ est une cog\`ebre fortement r\'eguli\`ere.
\\

Soit  $x :L \rightarrow Inv_{d,r}(F,F) $  une application lin\'eaire de la
cog\`ebre $L$
dans les op\'erateurs invariants \`a droite r\'eguliers sur la cog\`ebre $F$  .\\

Soit $U_{x}(\Delta_{L},\epsilon_{L})$  la big\`ebre donn\'ee par le th\'eor\`eme 2.2,
engendr\'ee par l'identit\'e de $T(F)$ dans $T(F)$ et les op\'erateurs $X(l) $ :

$$X(l) \in Inv_{d}(T(F),T(F) \subset Hom(T(F),T(F))$$
Hypoth\`ese:\\
Supposons qu'il existe une solution dans  $U_{x}(\Delta_{L},\epsilon_{L})$
au syst\`eme
d'\'equations:
$$\sum_{k} X(l_{k}^{'})\circ  Y(l_{k}^{''}) =\epsilon_{L}(l)1_{U_{x}}$$
$$\sum_{k} Y(l_{k}^{'})\circ  X(l_{k}^{''}) =\epsilon_{L}(l)1_{U_{x}}$$

o\`u
$$  \Delta _{L}(l) = \sum_{k} l_{k}^{'}\otimes  l_{k}^{''} $$
Alors la solution (dans $U_{x}$ ) est unique et v\'erifie

$$\Delta _{L}(Y(l)) =\sum _{k}Y(l_{k}^{''})\otimes Y(l_{k}^{'}) $$

Soit $T(L)(\Delta _{L} , \epsilon _{L})$ la big\`ebre universelle et soit $
I_{x}$
l'id\'eal associ\'e par le morphisme d'alg\`ebres :
$$ \pi  :  l_{i1}^{j1}\otimes l_{i2}^{j2}\otimes \ldots \otimes
l_{in}^{jn}\rightarrow
X(l_{i1}^{j1})\circ X(l_{i2}^{j2})\circ \ldots \circ X(l_{in}^{jn}) $$

Soit $ S_{1}$ de $ L \rightarrow T(L) $ une application lin\'eaire
v\'erifiant:\\

$ \pi \circ S_{1}(l) = Y(l)$ ,  et  soit : \\

$S^{r} :T(L) \rightarrow T(L)     $ l'extension de $S_{1} $ par
anti-homomorphisme ;

Alors :\\

A)\hspace{0.3cm}  l'id\'eal $I_{x}$  est un coid\'eal  de
$T(L)(\Delta_{L},\epsilon_{L})$ .\\

B) \hspace{0.3cm} Il existe un id\'eal minimal : $J_{x}$ ,
 de la big\`ebre libre  $T(L)(\Delta _{L} ,\epsilon _{L}
)$ , qui  v\'erifie:
\\

1)\hspace{0.3cm}  $I_{x}\subset  J_{x}$\hspace{0.3cm}, \\

2)\hspace{0.3cm}  $ J_{x}$ est laiss\'e invariant par $S^{r}$\hspace{0.3cm} ,\\

3)\hspace{0.3cm}  $ J_{x}$ est un coid\'eal de la big\`ebre  $T(L)(\Delta
_{L} ,\epsilon
_{L} )$
\hspace{0.3cm} .\\

C) \hspace{0.3cm}   $ H_{x} = T(L)(\Delta _{L},\epsilon _{L})/J_{x} $   est
l' alg\`ebre de
Hopf, canoniquement associ\'ee \`a la big\`ebre $U_{x}(\Delta _{L},\epsilon
_{L})$  .

\end{thm}

La d\'emonstration est essentiellement celle du th\'eor\`eme 3.2   .
L'int\'er\^et des th\'eor\`emes 3.1 et 3.2 est pr\'ecisemment de donner des conditions
qui permettent de montrer la validit\'e de l'hypoth\`ese faite dans le th\'eor\`eme 
4.1  .\\

\vspace{5cm}



\end{document}